\documentclass[%
 aip,
 amsmath,amssymb,
 reprint,%
nofootinbib]{revtex4-2}

\usepackage{graphicx}
\usepackage{float}
\usepackage{dcolumn}
\usepackage{bm}
\usepackage{comment}
\usepackage[utf8]{inputenc}
\usepackage[T1]{fontenc}
\usepackage{etoolbox}
\usepackage[plainpages=false,pdfpagelabels,colorlinks=true,linkcolor=black,urlcolor=black,citecolor=black,pdftitle={},pdfauthor={},pdfdisplaydoctitle=true,pdfduplex=DuplexFlipLongEdge,hypertexnames=true]{hyperref}

\newcommand{\beginsupplement}{%
        \setcounter{table}{0}
        \renewcommand{\thetable}{S\arabic{table}}%
        \setcounter{figure}{0}
        \renewcommand{\thefigure}{S\arabic{figure}}%
     }

\begin{document}

\makeatletter
\def\@email#1#2{%
 \endgroup
 \patchcmd{\titleblock@produce}
 {\frontmatter@RRAPformat}
 {\frontmatter@RRAPformat{\produce@RRAP{*#1\href{mailto:#2}{#2}}}\frontmatter@RRAPformat}
 {}{}
}%
\makeatother

\preprint{AIP/123-QED}

\title[]{In situ detection of RF breakdown on microfabricated surface ion traps}
\author{Joshua M. Wilson}

\author{Julia N. Tilles}
\author{Raymond A. Haltli}
\author{Eric Ou}
\author{Matthew G. Blain}
\author{Susan M. Clark}%

\author{Melissa C. Revelle}
\affiliation{ 
Sandia National Laboratories, Albuquerque, NM 87123 
}%

\date{\today}

\begin{abstract}
Microfabricated surface ion traps are a principle component of many ion-based quantum information science platforms. The operational parameters of these devices are pushed to the edge of their physical capabilities as the experiments strive for increasing performance. When the applied radio-frequency (RF) voltage is increased too much, the devices can experience damaging electric discharge events known as RF breakdown.  We introduce two novel techniques for in situ detection of RF breakdown, which we implemented while characterizing the breakdown threshold of surface ion traps produced at Sandia National Laboratories. In these traps, breakdown did not always occur immediately after increasing the RF voltage, but often minutes or even hours later. This result is surprising in the context of the suggested mechanisms for RF breakdown in vacuum. Additionally, the extent of visible damage caused by breakdown events increased with applied voltage.  To minimize the probability for damage when RF power is first applied to a device, our results strongly suggest that the voltage should be ramped up over the course of several hours and monitored for breakdown.
\end{abstract}

\maketitle

\begin{figure*}[t]
\includegraphics{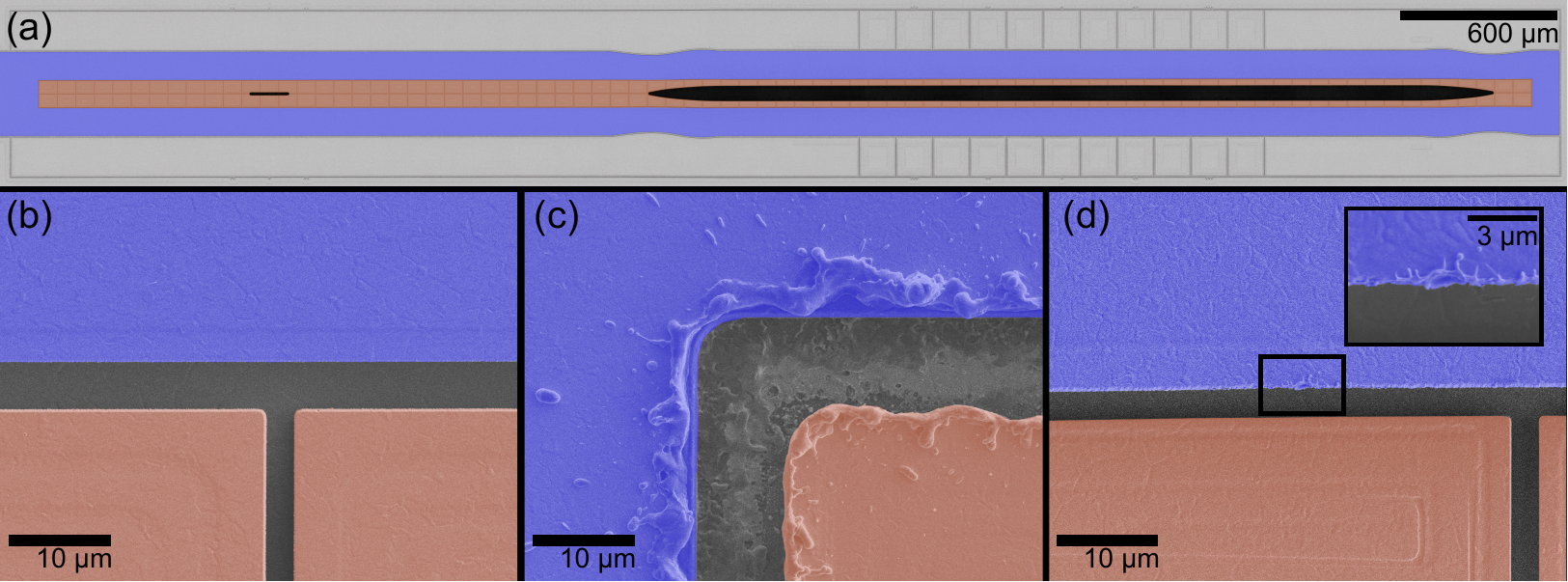}
\caption{Microscope images of several ion traps (false color added). In (a) there is an optical image of a Phoenix ion trap. The segmented, DC confinement electrodes (red) are surrounded by the RF electrode (blue).  The small black slot on the left is a dedicated loading hole, while the large black slot on the right can be used for loading or optical access to the ions. Images (b-d) are taken with an SEM. As in (a) the RF electrode is colored blue, and the DC confinement electrodes are colored red.  The grey area in between is a ground plane beneath the electrodes.  The device in (b) shows no breakdown induced surface damage. Image (c) shows a device which has had a sudden ramp in voltage ($\approx500$ V) and a breakdown event that led quickly to a permanent RF-ground short. The damage to the device covered a region an order of magnitude bigger than what is shown. The gap exposing the ground plane in (c) was originally the same size as the top gap in (b), but the breakdown event altered and deformed the overhanging electrode edge away from the gap. The device shown in (d) has slightly different geometry than the devices in (a) and (b), but the colors correspond to the same device regions.  This device had a single breakdown event at a much lower voltage ($\approx150$ V). The damage is barely visible at this scale. The inset to (d) is a magnified look at the region enclosed by the black rectangle.}\label{fig_SEM}
\end{figure*}

Trapped ions are a leading platform across many quantum information experiments including miniature atomic clocks \cite{lacroute_2018}, quantum sensors\cite{bollinger_2021}, and programmable quantum computers\cite{zoller_1995,wineland_2008,monroe_2013}. By manipulating thier internal states, trapped ions can be used as identical quantum bits (qubits) with very long coherence times, and arbitrary pairs of ions in a chain of ions can be entangled\cite{monroe_2019,landsman_2019}. A prominent potential platform for creating scalable trapped ion chains is microfabricated surface ion traps\cite{bruzewicz_2019,ospelkaus_2019,pino_2020,grzesiak_2020,clark_2021,nam_2020,blain_2021}. Such traps are often made using complementary metal-oxide-semiconductor (CMOS) compatible technologies which allow for high production repeatability of structures at the sub-micron level\cite{blain_2021}.  

In a surface ion trap, RF rails typically provide a radially confining pseudo-potential\cite{house_2008}. More RF power allows for higher secular frequencies and greater ion to surface-electrode separation, both of which reduce the magnitude of anomalous heating on the ions\cite{monroe_2006,wunderlich_2018}. The RF voltages used for surface trapping can range from a few tens of volts to several hundred, depending on the mass of the ion, the electrode configuration, and the desired trapping frequencies\cite{hong_2016,mcloughlin_2011}. Such voltages lead to very strong electric fields between surface electrodes, which have spacing ranging from one to ten micrometers for typical devices\cite{hong_2016,clark_2021,revelle_2020}.

When the voltage applied to a device is too high, discharge can occur between electrodes, a process known as RF breakdown\cite{raizer_1997,ospelkaus_2019, sterling_2013}. This discharge can damage the electrode metal, causing problems ranging from slightly distorted electric fields to physically shorted electrodes and total trap failure. Examples of damage from RF breakdown can be seen in the scanning electron microscope (SEM) images in Figure~\ref{fig_SEM}. Since not all breakdown events cause fatal trap damage, it is possible to have breakdown occurring for days or even months without any obvious external signs unless the device fails by developing a permanent short.  Even without device failure, such breakdown damage can contribute to trap instability which is difficult to distinguish from other possible noise sources. 

The nature and cause of RF breakdown in vacuum is an open area of research without complete agreement\cite{norem_2021}. As a fundamental part of many electronics applications, much effort has gone into understanding and mitigating RF breakdown. Unfortunately, varying theories and experiments have thus far yielded incomplete and often conflicting results\cite{juttner_2001}. This lack of theoretical understanding, coupled with difficulties of microfabrication, make it (currently) impossible to know a priori how much voltage a device will be able to sustain without breaking down. In this paper we present useful techniques for measuring breakdown in situ, as well as recommendations for safe use of microfabricated RF devices. Although it is beyond the scope of this paper to contribute to the theoretical discussion of mechanisms that induce RF breakdown in vacuum, we did find an unexplained delay in the time before breakdown which could help motivate future theoretical and experimental investigations.

Most of the surface ion traps utilized at Sandia are connected to routing circuits inside an ultrahigh vacuum chamber with optical access and appropriate electronic vacuum feedthroughs as described in Refs~\citenum{clark_2021} and \citenum{revelle_2020}. Figure~\ref{fig_SEM}(a) shows the RF and direct-current (DC) electrodes on a Phoenix trap (a linear, multi-layered surface ion trap fabricated at Sandia National Laboratories\cite{revelle_2020}).  We couple RF power to our devices with a helical resonator, which uses inductive coupling to both filter out all but a specific resonance frequency and increase the voltage applied at the device\cite{hensinger_2012}. Such a resonator system applies a root mean square voltage ($V_{rms}$) to the trap that can be estimated as\cite{hensinger_2012}:
\begin{equation}\label{eq_Vrms}
 V_{rms} = \frac{V_{peak}}{\sqrt{2}}\approx \kappa \sqrt{PQ},
\end{equation}
where $V_{peak}$ is the peak voltage, $P$ is the applied RF power, $Q$ is the quality factor, and 
\begin{equation}
 \kappa = \left(\frac{L}{C}\right)^{1/4},
\end{equation}
where $L$ and $C$ are the inductance and capacitance of the system respectively. This approximation relies on the assumption that the capacitance of the device dominates the capacitance of the combined system (including feedthroughs, wiring, etc.). As our fabrication techniques have improved and device capacitance has lowered, this assumption has become worse. When possible we calibrate the applied RF voltage using the secular RF frequency of a trapped ion and known trap dimensions. $Q$ is found by dividing the system resonance $f_0$ by the full width at half maximum $\delta f_{fwhm}$ of the reflected power spectrum\cite{hensinger_2012}. 

\begin{figure}
\includegraphics{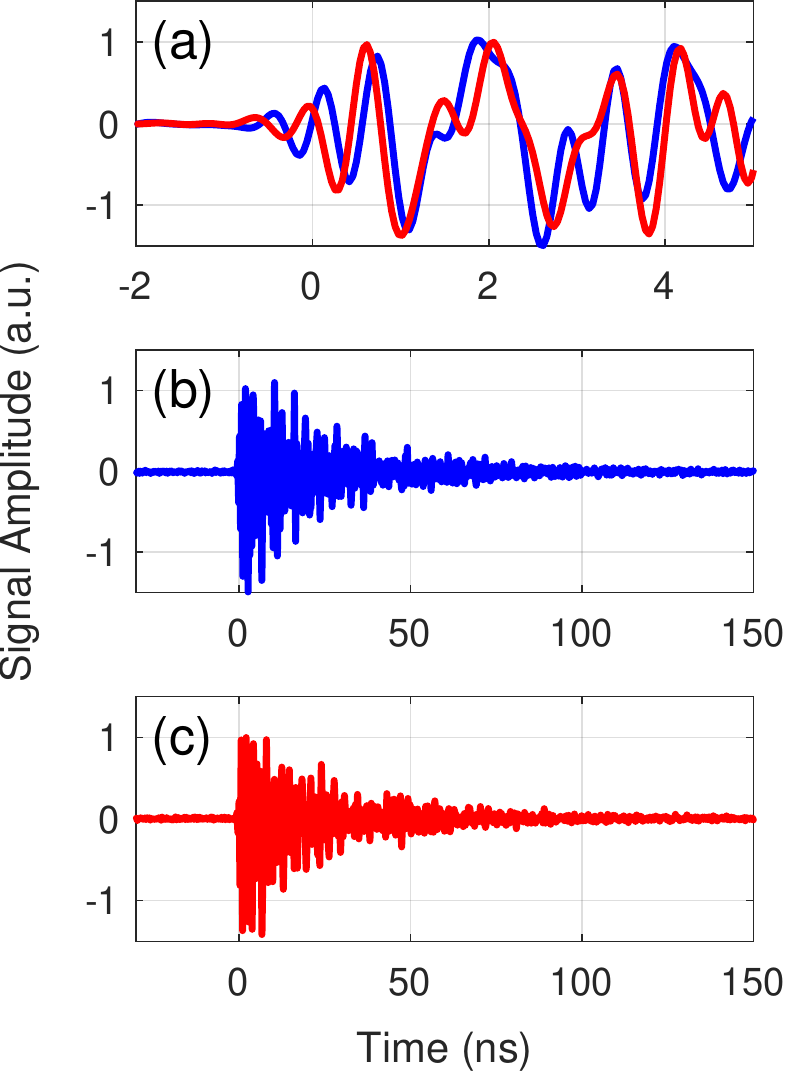}
\caption{Remote RF breakdown detection. Overlapped signals from two D-Dot detectors are seen in (a). Geometric considerations related to the detection axis of the D-Dots, and their placement relative to the trap, as well as signal reflections from the vacuum chamber and environment, prevent the traces from being identical. However, they are similar in structure for the first several nanoseconds, and their temporal coincidence shows that the detectors are nearly equidistant from the discharge source. In (b) and (c) the same two signals are shown at a longer scale so that the entire $\approx100$ ns pulse generated by RF breakdown is visible. The signal before and after the pulse is unstructured noise.}\label{fig_DDotTrig}
\end{figure}

While working to better understand and characterize breakdown of devices made at Sandia, we prepared a test chamber in which we could quickly exchange traps. We performed device installation in a cleanroom and then used a turbo pump to evacuate the chamber for at least 48 hours until the pressure was $< 3\times10^{-7}$ mbar. The chamber was oriented such that the surface of an installed device faced a window which was perpendicular to the table surface. On the outside of the chamber we placed D-Dot sensors (3.5 GHz bandwidth, free-space, electric field sensing devices by Prodyn\cite{ddot}), with a direct line of sight through the window to the trap. By amplifying and filtering their signal into a fast (>2 GHz) oscilloscope, we were able to detect electric field pulses generated by RF breakdown emanating from our ion trap devices (Figure~\ref{fig_DDotTrig}). The signals shown in Figure~\ref{fig_DDotTrig} were taken with D-Dots each set approximately one meter from the ion trap. The near coincidence of the signal on different D-Dots, as well as their similar shapes, made the source clearly distinguishable from other discharges that occurred regularly in the lab. This measurement is similar to work done at Sandia for lightning detection imaging, except we did not need to implement sophisticated cross-correlation (as done in Ref~\citenum{tilles_2019}) since the breakdown signals were so distinct. More details about the D-Dot measurement setup and execution can be found in the Supplementary Material.

We verified that this technique was registering single breakdown events by characterizing a device before and after a measured discharge. First by visually characterizing a device using an optical microscope at $500\times$ magnification and documenting the surface condition before beginning testing. After installing the device in our chamber we slowly increased the voltage until the D-Dots detected an apparent breakdown event. We then removed the device and performed another visual inspection for any sign of surface damage. With our optical microscope we saw a single new blemish and then further inspected that region with a scanning electron microscope (SEM). The result of this measurement can be seen in Figure~\ref{fig_SEM}(d). There is clear visual evidence of breakdown that was not present in the initial inspection of the device. 

\begin{figure}[t]
\includegraphics{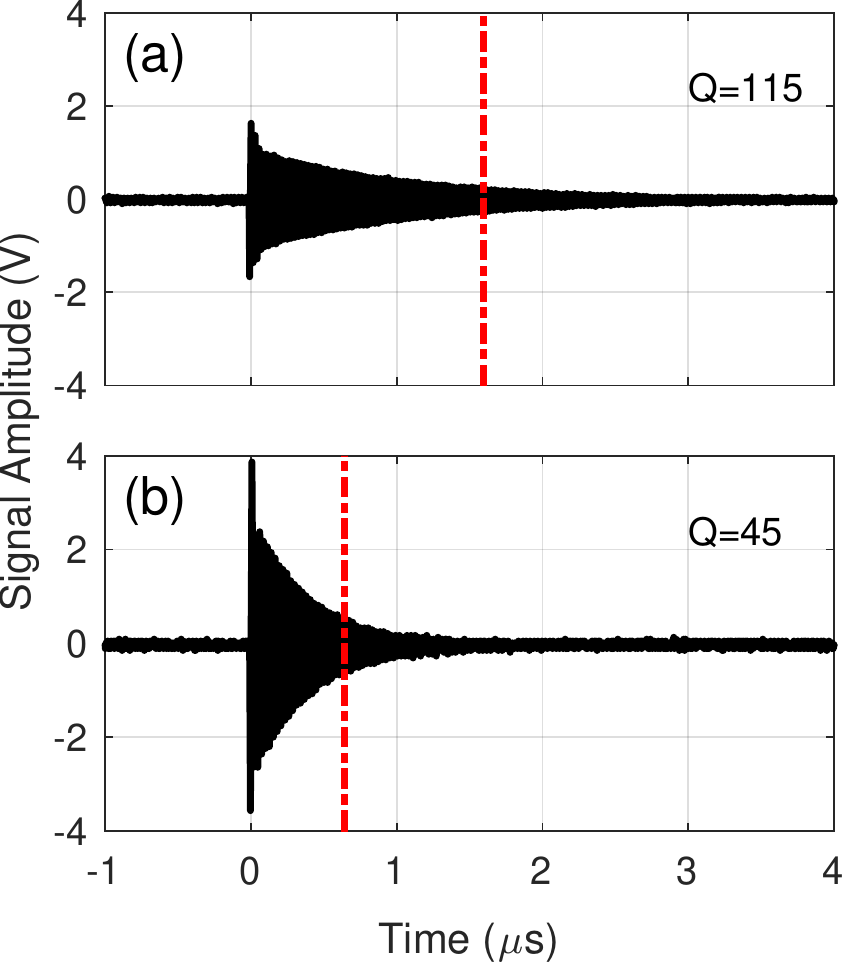}
\caption{Self-triggered signals of the RF back-reflection during two separate RF breakdown events. Each event was on a different device where (a) had a $Q$ of $115\pm3.4$ and (b) had a $Q$ of $45\pm1.3$. The black traces are the 35 MHz back-reflection spiking and then decaying. The spike is coincident with a D-Dot detection like the ones plotted in Figure~\ref{fig_DDotTrig}. The back-reflection takes a longer time to decay to its initial state after a breakdown than the RF signals picked up by the D-Dots. The time it takes each signal to decay to $1/e^2$ its maximum ($t_{decay}$) is plotted in (dashed) red. We find that $t_{decay}$ is proportional to $Q$, with a proportionality  $c_0=0.0140(3)\,\mu s$.}\label{fig_SelfTrig}
\end{figure}

Once establishing the success of this technique for in situ detection of even a very small breakdown event, we used the oscilloscope that was being triggered by the D-Dots to trigger a second scope. The second scope measured an RF signal reflected back from the resonator and through a directional coupler (see Figure~\ref{fig_sketch_BR} in the Supplementary Material). We aimed to find a reliable signal that would be identifiable on other experiments without needing the addition of D-Dot sensors. We discovered that breakdown events had a large and measurable effect on the back-reflected RF, making them easy to detect with a simple oscilloscope. Figure~\ref{fig_SelfTrig} shows the RF back-reflection during a breakdown event on two different devices. The signals shown were observed simultaneously with D-Dot detection. Although the breakdown event (Figure~\ref{fig_DDotTrig}) and the interruption to the forward power on the device is short, the effect on the back-reflection takes a relatively long time to decay. The signal is easily measurable on any oscilloscope with larger bandwidth than the RF signal being applied, and could be used as its own trigger source. The inductance of the resonator circuit prevents a quick return to equilibrium after breakdown occurs. On average, the time the signal took to decay to $1/e^2$ of its maximum ($t_{decay}$) was:
\begin{equation}
    t_{decay}=c_0 Q
\end{equation}
with $c_0$ equal to $0.0140(3)\,\mu s$.  This was only tested in one chamber, but with two different devices.  Each breakdown event observed on the same device had the same measured $t_{decay}$ , but we do not know if $c_0$ would be the same in other experiments. For more information about this breakdown measurement technique please see the Supplementary Material.

We had seen several devices suffer breakdown events, which led to a short from RF to ground, many hours or even days after applying a specific voltage. Because of this prior experience, instead of continuously turning up the voltage until observing instantaneous breakdown, we increased the voltage in steps of a few percent relative to the expected breakdown voltages. After increasing the voltage we waited for at least 40 minutes before moving to the next voltage. (Preliminary measurements suggested that 20 minutes would be sufficient to observe most breakdown events, and waiting for days at every voltage step was too costly. Some measurements ended up being much longer.)

As the voltage was increased the chamber pressure would also increase. Starting pressures during our measurements ranged from a little below $10^{-10}$ to $\approx3\times10^{-7}$ mbar. We paused measurements and allowed more time for pumping if the pressure exceeded $9\times10^{-7}$ mbar. We would resume measurements from a lower starting pressure after a day or two. Though the pressures were higher than the ultrahigh vacuum used when trapping ions, the extra week that would have been required to bake and prepare ultrahigh vacuum would have greatly reduced the number of traps we tested. Some devices were tested with different starting pressures (including one at $<10^{-10}$ mbar) to check the dependence of breakdown voltage on pressure. We found no correlation at the scale of other uncertainties for pressures below $10^{-6}$ mbar.

We tested several device geometries where the minimum distance from the RF electrode to ground ranged from 1 to 5 $\mu$m. However, the details of the fabrication procedures, and the long delay before breakdown, led to a large amount of variability in the voltage at which we first measured breakdown for a specific device. We were unable to find a clear correlation between the gap size and the breakdown voltage.  We also found that breakdown would occur at unexpectedly low voltages in some devices, but it was isolated and we could increase the voltage before breakdown became more regular. This was likely due to an asperity under the RF electrode, which caused a local enhancement in the electric field. After a discharge burned off the asperity, higher voltages were sustainable.

\begin{figure}
\includegraphics{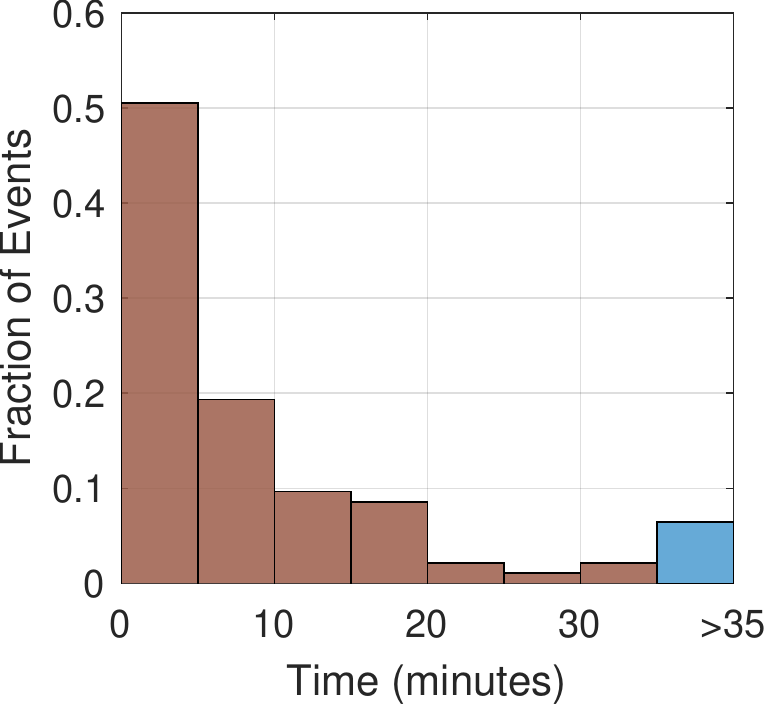}
\caption{Distribution of minutes before breakdown across multiple tested devices. This histogram shows the results of 93 distinct breakdown events. We recorded the time at which a breakdown event occurred after a voltage change, or the delay between separate events at the same voltage. Roughly half of the breakdown events happened within five minutes, and the majority happened within 20 minutes. Four events occurred at an interval of more than 35 minutes, with the longest being almost 200 minutes. Not every voltage was held for that long, but all measurements were held for at least 40 minutes at each voltage. The temperature and the pressure in the trap should be stable well within that time.}\label{fig_Histogram}
\end{figure}

Another approach which has been used for measuring breakdown, introduced in Ref.\citenum{sterling_2013}, is to monitor the surface of the device using a camera. In this case, the voltage is continuously increased until an arc occurs and is observable. This approach requires breakdown that is large and optically bright enough to be visible on the device surface.  Continuously increasing the voltage for such a measurement makes sense under the assumption that a device either can or cannot sustain a particular voltage and that past some threshold breakdown is inevitable and sudden. Other RF breakdown measurements and theories seem to use similar assumptions\cite{juttner_2001,norem_2005,norem_2021}. In our measurement, however, we observed that a trap could be kept at a specific voltage for a long time before a breakdown event would occur. We saw events occur as quickly as a few seconds and up to several hours after we changed to a new voltage, as seen in the histogram in Figure~\ref{fig_Histogram}. 

We also found that breakdown events occurring at lower voltages caused qualitatively less damage than those at higher voltages. In fact, Figure~\ref{fig_SEM}(c) shows the damage when a device was accidentally ramped to a higher than desired voltage and caused a nearly instant breakdown and device short. The damage to the device was more consistent with the instantaneous breakdown measurements in Ref.~\citenum{sterling_2013} than with our other, more localized events at lower voltages (Figure~\ref{fig_SEM}(d)). Furthermore, we don't know if the discharge causing trap damage is a "cold" breakdown like coronal discharge (which wouldn't necessarily be detected by a camera), or something "hot" and highly luminous like arcing\cite{raizer_1997} (or perhaps some combination of discharge types).  A wide range of breakdown types (if they occur) would be indistinguishable with our setup.

These results highlight the difficulty of predicting when and at what RF voltage breakdown will occur in micro-fabricated devices at high vacuum. On our devices, breakdown may occur minutes or hours after the voltage has been set, suggesting complex breakdown physics. Hopefully this work can motivate new directions in the theoretical efforts to understand RF breakdown in vacuum. 

Our results show that great care should be taken when the RF voltage for microfabricated surface ion traps is being set. Voltages should be ramped slowly to minimize damage during isolated breakdown events. Low voltage breakdown can possibly remove problematic regions of the trap where breakdown is more likely to occur. In addition, the back-reflection should be monitored for breakdown. Signals like the one in Figure~\ref{fig_SelfTrig} should be straightforward to measure in other ion trapping experiments. If multiple events are observed at or near the same voltage, the RF power should be immediately reduced to avoid the development of permanent shorts.

Despite the utmost care, permanent trap damage is still possible. There is wide variation in the behavior of an RF breakdown event, and in some of our tests, the first observed breakdown caused a permanent short. This only ever occurred when we were pushing devices above the level recommended for safe operation on a specific device design. The techniques outlined in this paper can help avoid many problems, but do not prevent them entirely.

The use of D-Dots to detect lab discharge events could be widely applicable as it does not require a particular signal source. Many sources of breakdown events could be detected, even events that are underneath some dielectric or structure. Such detection in combination with the cross correlation techniques described in Ref. \citenum{tilles_2019} can allow for pinpointing the source location of a breakdown in a lab if needed. As we demonstrated the signal detected by D-Dots could also be used to trigger measurements of other signals or cameras.

Moving forward we will use the techniques demonstrated here to avoid breakdown-induced damage to our existing ion trap experiments. We will also investigate other means to engineer traps that are less susceptible to RF breakdown. 

\begin{acknowledgments}
This material was funded in part by the Office of the Director of National Intelligence (ODNI), Intelligence Advanced Research Projects Activity (IARPA) under the Logical Qubits (LogiQ) program.
It was also supported in part by the U.S. Department of Energy, Office of Science, Office of Advanced Scientific Computing Research Quantum Testbed Program.

Sandia National Laboratories is a multimission laboratory managed and operated by National Technology \& Engineering Solutions of Sandia, LLC, a wholly owned subsidiary of Honeywell International Inc., for the U.S. Department of Energy’s National Nuclear Security Administration under contract DE-NA0003525. This paper describes objective technical results and analysis. Any subjective views or opinions that might be expressed in the paper do not necessarily represent the views of the U.S. Department of Energy or the United States Government.
\end{acknowledgments}

\section*{Data Availability Statement}
The data that support the findings of this study are available from the corresponding author upon reasonable request.

\section*{Supplementary Material}
Supplementary Material can be found online at:(below for now)
\bibliographystyle{aip}
\bibliography{BreakdownRefs.bib}
\pagebreak

\onecolumngrid
\section*{Supplementary Material}
\beginsupplement

\setcounter{page}{1}

\section{D-Dot measurement details}\label{SM_ddot}
\renewcommand{\thefigure}{S\arabic{figure}}
\setcounter{figure}{0}
\begin{figure}[h!]
\includegraphics[width=0.85\textwidth]{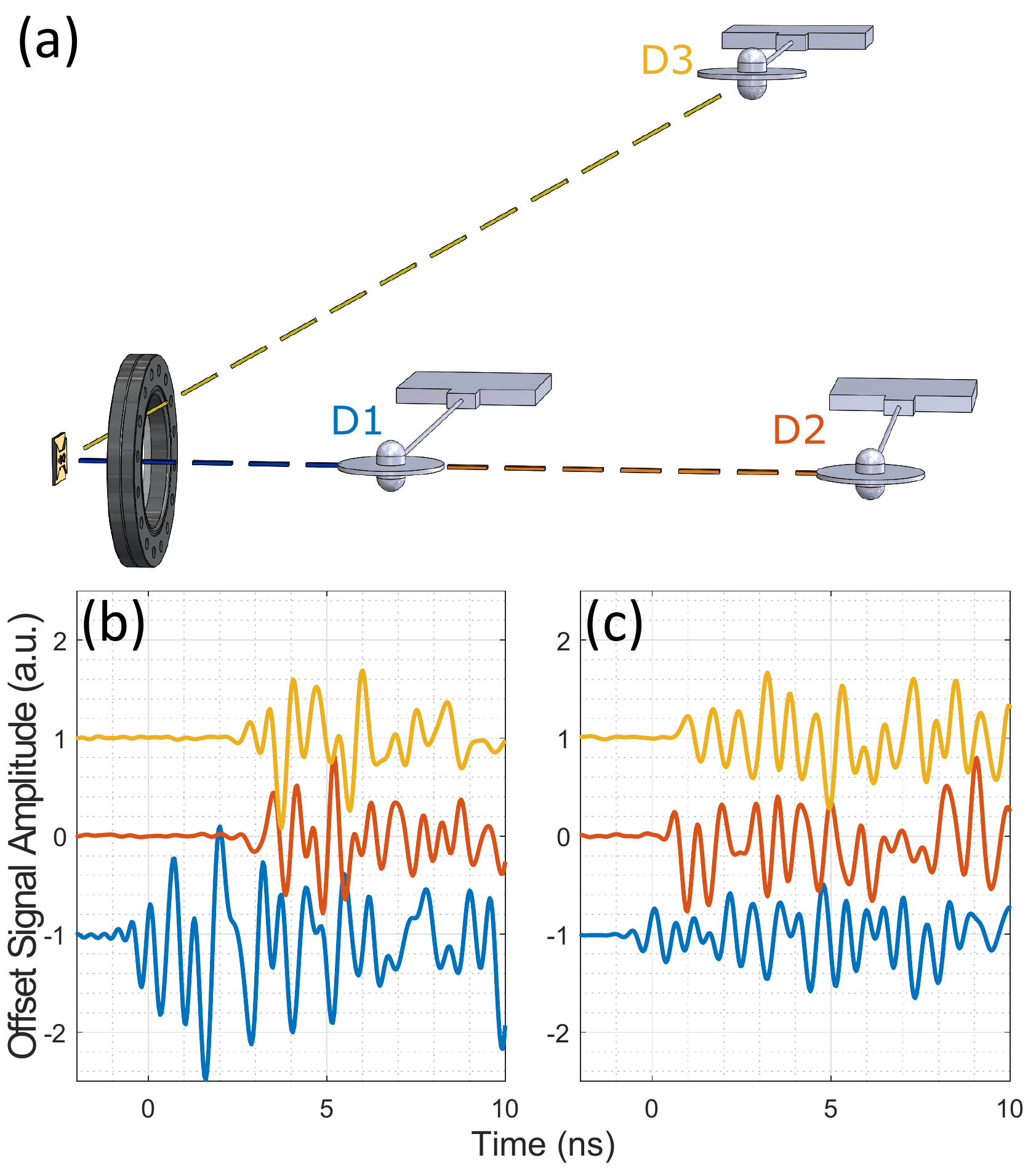}
\caption{Details of the D-Dot measurement. The sketch in (a) shows the RF driven trapping device (yellow bow-tie shape) inside the vacuum chamber (only window is shown).  A breakdown event emits an RF pulse that is detected by detectors D1, D2 and D3.  The disk shape on the front of each detector is the D-Dot, and the slim box shape on the rear is the balun.  In the setup as displayed D2 and D3 are equidistant from the trapping device, so the yellow dashed line is the same length as the blue and orange lines combined.  The D-Dot measured breakdown signal in (b) arrived at D1 about 3 ns before it arrived at D2 and D3.  This time delay corresponds with speed of light transit across the physical separation of the detectors.  D2 and D3 were one meter farther from the source than D1.  In (c) there is no such delay, although the detector setup was the same.  Thus it was clearly generated by some other lab discharge.  In (b) and (c) the signals are offset vertically for clarity.  In reality they all were centered around zero.  Also, only the first 10 ns of a signal that lasted much longer is shown (see Figure~\ref{fig_DDotTrig}).}\label{fig_sketch_Ddot}
\end{figure}
The D-Dots and corresponding equipment were set-up external to the vacuum chamber, allowing for the assembly to occur around a live experiment. The D-Dots themselves were  Prodyn AD-70s\footnote{\url{https://www.prodyntech.com/wp-content/uploads/2021/01/AD-Series-Free-Field-D-Dot-Data-Sheet.jpg}} (sketched as a grey disk in Figure~\ref{fig_sketch_Ddot}) which operate from DC to 3.5 GHz.  Each D-Dot is connected to a Prodyn BIB-100G balun\footnote{\url{http://www.prodyntech.com/wp-content/uploads/2013/09/Balun-Spec-Sheet-01292018212240.pdf}} which operates from 250 kHz to 10 GHz.  The balun (sketched as a slim grey box in Figure~\ref{fig_sketch_Ddot}) takes a differential measurement from each side of the D-Dot and outputs a single value. We then amplified the signal (Pasternack PE15A3311: 0.1 MHz to 10 GHz, 32.5 dB gain) and measured it with a fast (>2 GHz) oscilloscope. To avoid noise and loss we used Pasternack PE-P142LL cables (low-loss triple shielded coax) for sending the signal to the scope.  

All of our measurements were conducted with 3 D-Dots.  We arranged the D-Dots in two main configurations during the measurement process.  Our initial measurements and calibration were done with an arrangement like the one sketched in Figure~\ref{fig_sketch_Ddot}(a).  D1 was placed about 30 cm from the trapping device.  D2 and D3 were about a meter farther from the device, but at different heights relative to each other.  D1 and D2 were at the same relative height as the trapping device.  The measured breakdown signals shown in Figure~\ref{fig_sketch_Ddot}(b) came from an event at the device and the signal in \ref{fig_sketch_Ddot}(c) did not.  This is clear because in \ref{fig_sketch_Ddot}(b) the signal from D1 proceeds the signals on D2 and D3 by about 3 ns.  It takes 3 ns for light to travel the extra meter past D1 to the other detectors.  

About half the measurements were performed with the D-Dots in a slightly different configuration.  All three D-Dots were placed equidistant from the trapping device, but with different angles relative to the device and the chamber.  In those measurements we looked for the signals to be coincident in time to distinguish them from other discharge events in the lab.  In both detector orientations trapping device based breakdown events had very similar shapes, which also helped us distinguish them from other signals that we detected.  

\section{Back-reflected signal measurement}\label{SM_backreflection}
\begin{figure}[h]
\includegraphics[width=0.6\textwidth]{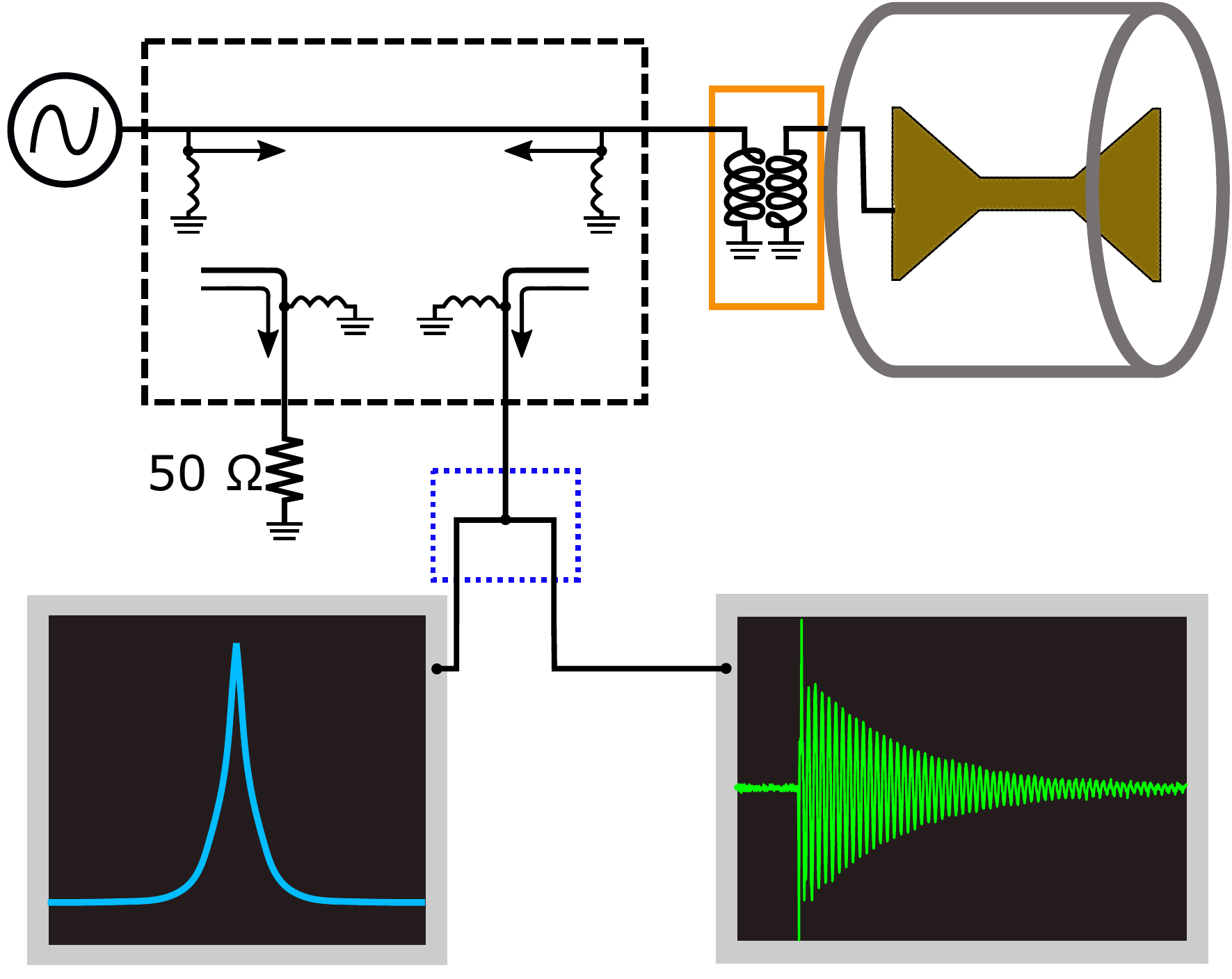}
\caption{Sketch of setup for measuring the breakdown signal in the back-reflected RF. The RF signal is sent through a bi-directional coupler (black dashed box), a helical-resonator (orange solid box), and through a vacuum feed-through into the vacuum chamber (grey cylinder).  In the vacuum chamber, the RF signal is connected to the ion trapping device (brown bow-tie).  The back-reflected signal is monitored and minimized (blue power spectrum).  $50~\Omega$ termination is essential on the unused coupler port.  When monitoring for breakdown, a power splitter (blue dotted box) sends half the the signal to an oscilloscope which uses the back-reflection as a trigger source.  Breakdown events in our setup generated signals like the green waveform (see also Figure~\ref{fig_SelfTrig}).}\label{fig_sketch_BR}
\end{figure}
Most of the ion-trapping experiments at Sandia National Labs have RF power delivered to the trapping device in accordance with the sketch in Figure~\ref{fig_sketch_BR}.  A low noise, pre-amplified RF signal goes through a bi-directional coupler (Mini-Circuits ZFBDC20-61HP-S+, 1-60MHz), a helical resonator, and a vacuum feed-through to get to the trapping device.  The back-reflected signal coming from the coupler is minimized to ensure optimal power delivery to the trapping device.  When optimizing, we typically measure a peak back-reflected signal amplitude of around 10 dBm, which is minimized to between -30 and -60 dBm.     

The only difference to the setup when monitoring for breakdown is adding a power splitter (Mini-Circuits ZFSC-2-5-S+, 10-1500 MHz) on the back-reflected path.  One half of the split signal is used for normal power coupling optimization.  The other half is sent to an oscilloscope in normal trigger mode using the input signal as a trigger source. As is visible in the green signal in Figure~\ref{fig_sketch_BR} (as well as in Figure~\ref{fig_SelfTrig}) the breakdown signals that we observed were much bigger than the 35 MHz background amplitude.  Setting the trigger at two to three times the background level was enough to avoid (most) spurious triggers and to capture a breakdown event when it happened.  The time scale that we observed was $Q$ times about 0.014 microseconds.  We don't know if this time scale will be consistent across different experiments.

\end{document}